\documentclass[prb,aps,twocolumn,superscriptaddress,groupaddress]{revtex4-1} 
\usepackage{amsmath}  
\usepackage{amsfonts} 
\usepackage{graphicx} 
\usepackage{tikz}
\usepackage[siunitx]{circuitikz} 
\usetikzlibrary{shapes,arrows}
\usetikzlibrary{decorations.pathmorphing}
\usepackage{siunitx}
\usepackage{wasysym}
\usepackage{booktabs}

\begin{document}

\title{A gaseous proportional counter built from a conventional aluminium beverage can}
\affiliation{Division of Elementary Particle Physics, Department of Physics, University of
Helsinki, Finland}
\affiliation{Helsinki Institute of Physics, Finland}

\author{Alexander Winkler}
\affiliation{Division of Material Physics Physics, Department of Physics, University of
Helsinki, Finland}
\affiliation{Helsinki Institute of Physics, Finland}
\author{Aneliya Karadzhinova}
\affiliation{Division of Elementary Particle Physics, Department of Physics, University of
Helsinki, Finland}
\affiliation{Helsinki Institute of Physics, Finland}
\author{Timo Hild\'{e}n}
\affiliation{Division of Elementary Particle Physics, Department of Physics, University of
Helsinki, Finland}
\affiliation{Helsinki Institute of Physics, Finland}
\author{Francisco Garcia}
\affiliation{Helsinki Institute of Physics, Finland}
\author{Giacomo Fedi}
\affiliation{Division of Elementary Particle Physics, Department of Physics, University of
Helsinki, Finland}
\author{Francesco Devoto}
\affiliation{Division of Elementary Particle Physics, Department of Physics, University of
Helsinki, Finland}
\affiliation{Helsinki Institute of Physics, Finland}
\author{Erik J. Br\"ucken}\email{erik.brucken@iki.fi}
\affiliation{Division of Elementary Particle Physics, Department of Physics, University of
Helsinki, Finland}
\affiliation{Helsinki Institute of Physics, Finland}
\email{erik.brucken@iki.fi}
\date{\today}

\begin{abstract}
The gaseous proportional counter is a device that can be used to detect
ionizing radiation. These devices can be as simple as a cylindrical cathode and a very thin anode wire centered along its axis. By applying a high voltage, a strong electric field is generated close to the anode wire. Ion-pairs, generated by passing ionizing radiation, create avalanches once they drift into the strong electric field region near the anode. The electrical charges created by the avalanche generate an observable signal which is proportional to the energy loss of the incoming radiation. We discuss the construction of such a device. Our detector was built from an ordinary aluminium beverage can and uses a common electric wire strand as the anode.  The construction of this detector offers students at universities or technically oriented high schools a detailed understanding of the design and operation of gaseous radiation detectors. The equipment required to complete the project should be available at most institutions.
\end{abstract}

\maketitle

\section{Introduction} 

In student laboratories there is a constant demand for simple and cost-effective  laboratory equipment to illustrate modern physics phenomena. In the area of nuclear or particle physics, radiation detectors can be quite expensive. Commercial detectors also do not offer as extensive learning factors as the ``do it yourself'' equipment does. In this article we discuss how a gaseous proportional counter can be built and used to acquire $X$- and $\gamma$-ray spectra from radioactive sources.

\begin{figure}[h]\centering
\includegraphics[width=0.4\textwidth]{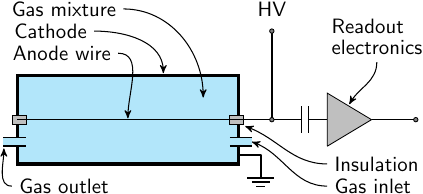}
\caption{Schematic drawing of a proportional counter.\label{Fig01}}
\end{figure}

\begin{figure}[h]\centering
\includegraphics[width=0.46\textwidth]{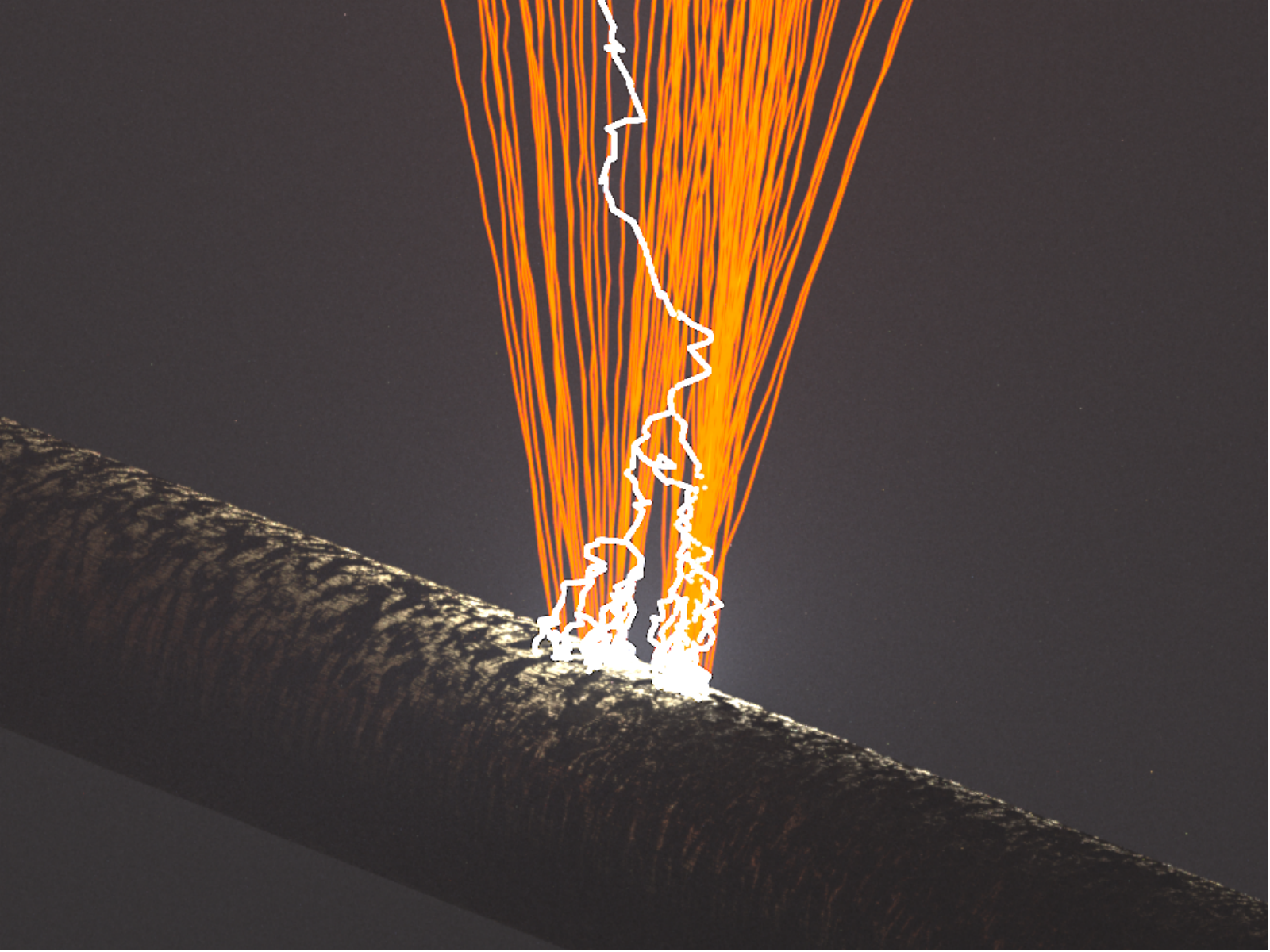}
\caption{Simulation of an avalanche caused by a single ionizing electron (white) in a proportional counter. The positive ions (orange) and part of the anode wire are shown as well.}\label{Fig02}
\end{figure}

A gaseous proportional counter \cite{rutherfordgeiger} operates with the same principle as an ionization chamber \cite{knoll} with the additional feature of gas amplification. A schematic drawing of a simple proportional counter is shown in Fig.\,\ref{Fig01}.
The radial electric field between the cathode and the anode is generated by applying positive high voltage (HV) to the anode. A photon or a charged particle that enters the chamber, ionizes the gas and the electrons (ions) move towards to the anode (cathode) and  induce an electric current. The electric field strength in a cylindrical detector can be approximated by
\begin{equation}
E(r)=\frac{U}{r\ln(b/a)},
\end{equation}
where $U$ is the applied potential, $b$ is the inner radius of the cathode tube, and $a$ is the outer radius of the anode wire. If the applied voltage is high enough, a sufficiently strong electric field of around $100$~kV/cm will be generated close to the anode wire and secondary ionization and Townsend avalanches will occur in this region.\cite{knoll} This process is commonly referred to as gas amplification. 
Figure~\ref{Fig02} illustrates this process. It shows the result of a simulation of our detector using \textsc{Garfield$^{++}$},\cite{garfield} a toolkit for simulation of gas and semi-conductor based particle detectors. The path of the drifting electron that results from the ionizing process is shown in white. Close to the anode ($\approx 30\,\mathrm{\mu m}$) the Townsend avalanche ignites. The positive ions that drift towards the cathode are shown in orange.
The term ``proportional'' indicates that the total number of electron-ion pairs is proportional to the number of primary electron-ion pairs created by the incident radiation. 
The energy of the incident radiation can only be measured if exactly one avalanche occurs per primary electron-ion pair.

A special gas mixture, mainly consisting of a noble gas, is used for optimal performance. Noble gasses can be excited and emit photons that can release electrons from the detector material through the photoelectric effect. These electrons cause additional avalanches that distort the primary signal.  To suppress such spurious pulses, some molecular anti-quenching gas, such as CO$_2$ or methane, is added. 

The avalanche electrons drift to the anode rather quickly ($<1$~ns) compared to the much slower positive ions ($>1$~$\mu$s). This leads to a fast rise of the signal caused by the electrons and a much slower rise caused by the ions. The amount of the induced charge by the electrons is very small compared to the amount induced by the ions.\cite{fsauli} 

If a very high voltage is applied to the proportional counter it operates as a 
Geiger-M\"uller counter. In this regime, each primary electron-ion pair will not only create an avalanche but also high-energetic photons that are emitted by excited atoms. This causes a full electric discharge that can be easily processed with simple electronics. However, the proportional character of the detector signal is lost.

For a more detailed introduction into gaseous proportional counters we refer the reader to the literature\cite{knoll,sauli} and references therein.

This article is organized as follows. Section~\ref{sec:assembly} describes the assembly of the proportional counter. In Section~\ref{sec:setup}, the experimental setup and the preparations and testing procedures are discussed. Section~\ref{sec:electrocalib} describes the calibration procedure. In Section~\ref{sec:spectra}, the spectra obtained with two radioactive sources are presented. Finally, the operation region of the proportional counter is probed and discussed in Section~\ref{sec:hvscan}. Our conclusions are presented in Section~\ref{sec:conclusions}.

This article has two educational components. The first component is the mechanical construction and setup of the detector and the electronics, described in Sections~\ref{sec:assembly}-\ref{sec:electrocalib}. The second component is the analysis of spectral data recorded with this detector, covered in Sections~\ref{sec:spectra} and \ref{sec:hvscan}. With this publication, we specifically concentrate on the detector itself. Operational equipment, such as the HV-supply, the oscilloscope, and the amplifier, are assumed to be available to the reader. These devices can be of the ``do it yourself'' type as well, but a detailed description of these would go beyond the scope of this publication. The interested reader is referred to publications\cite{lowCostMCA, lowCostPreAmp, lowCostHVDC} or websites specialized on these  topics.

\section{Detector assembly}\label{sec:assembly}

The motivation for a radiation detector made of an aluminium beverage can is to build a properly working device at minimal costs. Similar approaches have been proposed in the past\cite{SewingNeedle, CounterOld} but have become partially impracticable due to the availability of suggested parts and the crude data quality these devices offer. Our approach follows the  design of Moljk and Drever,\cite{CounterOld} but uses materials that are available in today's society. In addition, our design significantly improves the performance of the device. 

\begin{figure}[]\centering
\includegraphics[width=0.45\textwidth]{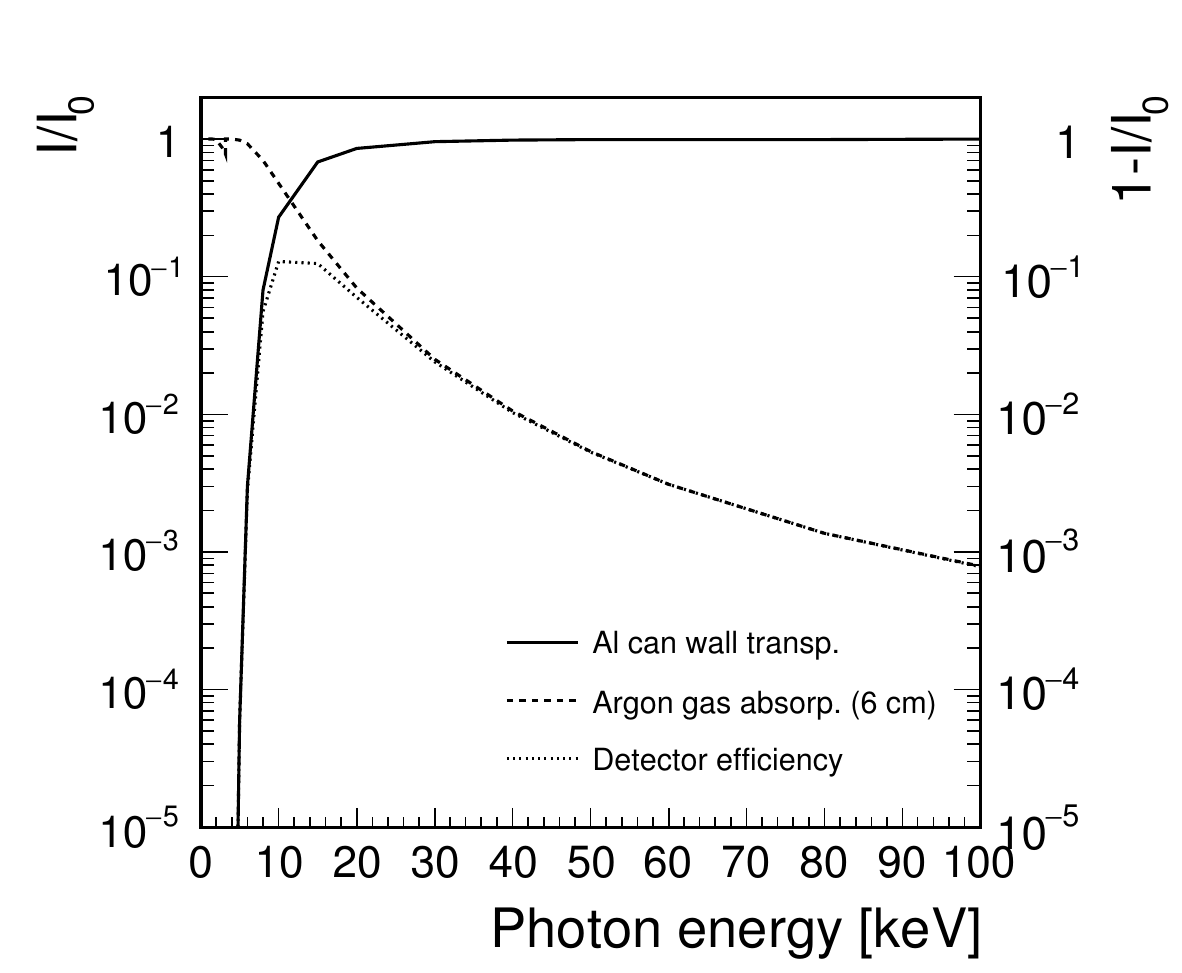}
\caption{The relative photon intensity versus incident energy due to absorption in the thin aluminium beverage can wall (isolid curve) and absorption in the 6 cm thick argon gas (dashed curve). The estimated detector efficiency is shown by the dotted curve. The plot was produced with data taken from the NIST Standard Reference Database.\cite{nist}}\label{Fig03}
\end{figure}

A good option for a cheap conducting cathode tube is a conventional aluminium beverage can.\cite{cansize} The wall thickness is thin enough for low energetic $\gamma$- and $X$-rays to enter the gas volume and no additional \textit{entrance-window}, as used by Moljk and Drever,\cite{CounterOld} is needed for operation. 
This is illustrated in Fig.\,\ref{Fig03} 
which shows the ratio of  the transmitted photon intensity $I$ after passing through detector materials and the incident photon intensity $I_0$ as function of photon energy. The black curve shows the results of absorption in the  thin aluminium can wall and the red curve the absorption in 6 cm of argon gas. Multiplying these two curves gives an rough estimate of the detector efficiency.
The photon energies of interest are between 5 and 60 keV. These energies are high enough to enter the detector through the can wall and low enough for some of the photons to be absorbed in the gas.

The parts used for our assembly are listed in Table~\ref{Tab01}. Our detector was designed and built from scratch without the use of prefabricated components. The beverage can was carefully opened from the top side with a hole saw. A small area on the outside of the can was roughened with sandpaper to expose the bare aluminium. It is recommended to remove the polymer coating inside the can for better detector performance. However, in our case the coating was not removed. Acrylglass covers were used as end-caps of the can. Holes for the anode wire and the gas supply were drilled in the bottom side of the can and in the Acrylglass covers. The diameter of the holes were chosen to be of the size of the nylon screws, M5 for the anode holes and M8 for the gas supply holes. Another hole was drilled into the can wall close to the opening to attach a firm ground connection for the cathode cable.
The central cores of the the nylon screws were removed to hold the brass tubes that support the anode wire and the gas supply lines. The diameters of the holes drilled into the M5 and M8 screws were 1.2~mm and 3~mm, respectively.  All components were cleaned using ethanol in an ultrasonic bath for about 3 minutes. 
\begin{table}[]\centering
\caption{Parts used to build the proportional counter 
described in this paper.
\label{Tab01}}
\begin{tabular}{lc} 
\hline\hline
Amount & Part \rule{0pt}{2.6ex} \rule[-1.2ex]{0pt}{0pt}\\ 
\hline
1 & Aluminium beverage can \rule{0pt}{2.6ex}\\
& (empty, length 168~mm, \diameter 67~mm,\\
& wall thickness 0.19~mm)\\
1 & Low voltage cable\\
& strand (\diameter $\approx 0.08$\:mm)\\
1 & High voltage (HV-) connector\\
1 & Low voltage cable (\diameter $\approx 2$\:mm)\\
& to connect cathode to the ground\\
& of the HV-connector\\
2 & Nylon screws and nuts (M5 size)\\
2 & Pieces of Acrylglass (\diameter $\geq$ bottom of can)\\
1 set & Two part epoxy adhesives\\
2 & Nylon screws and nuts (M8 size)\\
4 & Metal rods (threaded, $>$ length of can)\\
12 & Nuts (fitting the metal rods)\\
1 & Mini banana plug and socket\\
& with screw hole\\
1 & Piece of rubber sheet (\diameter of can)\\
1 & Strip of copper tape\\ 
2 $\times$ 4\:cm&Brass tube (inner \diameter $\lesssim 1$\:mm) to hold the anode\rule[-1.2ex]{0pt}{0pt}\\
\hline\hline
\end{tabular}
\end{table}

\begin{figure*}[]\centering
\includegraphics[width=0.98\textwidth]{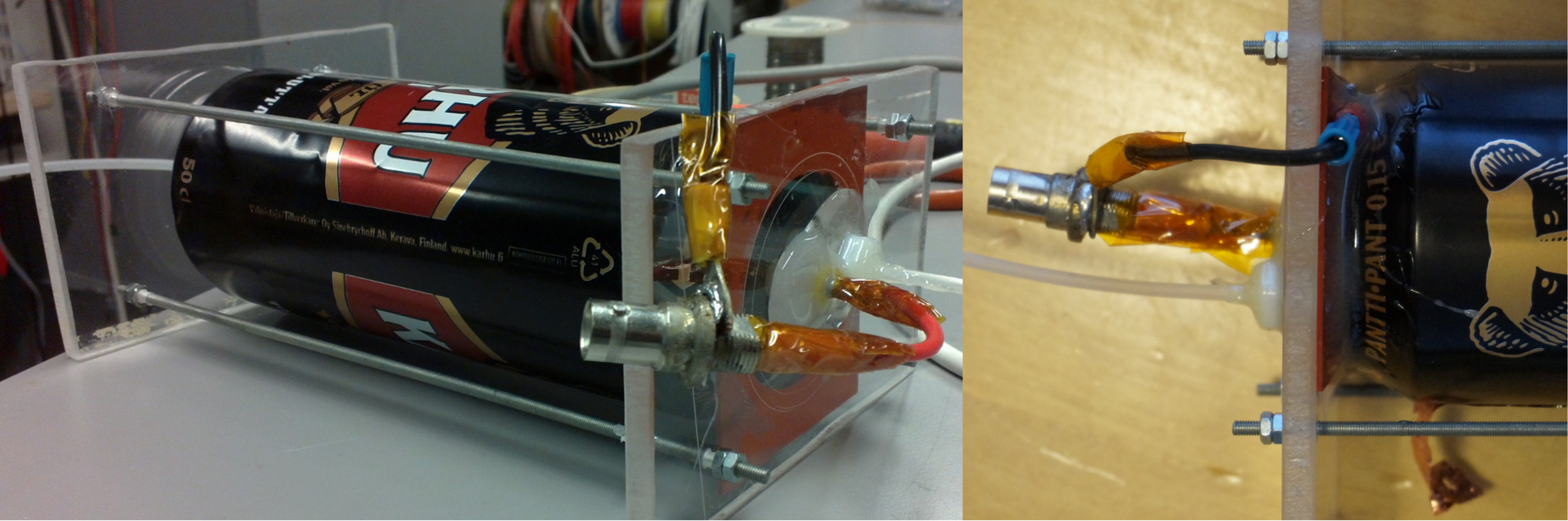}
\caption{A gaseous proportional counter built from an aluminium beverage can.}
\label{Fig04}
\end{figure*}

The components were assembled as shown in Figs.~\ref{Fig01} and \ref{Fig04}. The nylon screws used to hold the gas supply lines were put in place first. Another nylon screw to hold the anode wire was fixed to the bottom of the can. This provides electrical insulation for the brass tube that supports the anode wire and 
keeps the anode wire centered inside the can. Our experience has shown that it is best to first put the anode wire carefully through the brass tube and then solder the brass tube to the HV-connector.
The thin anode wire was obtained by taking a single strand from a conventional head-phone cable. The drawback of this approach is a larger wire diameter and uneven fabrication compared to e.g. professional beryllium-copper wires with a diameter of 50~$\mu$m or less. The effect of the thicker wire can be compensated with higher operational voltages. 
The can was closed and tightened after pulling the wire carefully through the brass tube at the opposite end. Rubber was used between the can and the acrylic end-caps. Since the rubber might not be sufficiently gas-tight, two part epoxy adhesives was applied to the end-caps to ensure gas-tightness.  The detector was gas tight 24~h after the application of the epoxy. 

The anode wire was soldered to the brass tube on the bottom of the can after straightening it carefully and the screw-on socket for the cathode cable was installed.
The gas supply tubes were connected and all possible areas of gas leakage were sealed with epoxy adhesive. 
To connect the cathode, a low voltage cable, $\approx 2$\:mm \diameter, was soldered to the metal casing of the HV-connector. A mini banana plug was attached to the other end of the cable to connect it to the socket attached to the can wall.
These cables are visible in the pictures in Fig.~\ref{Fig04}.  In the left picture the red high voltage connection is visible on the right-hand side. On the left-hand side of the right picture, the black cable providing the ground connection to the cathode is visible. Discharge can occur if a non-HV rated cable is used for the anode connection since most non-HV rated cables are only rated up to 500~V and a discharge between HV and ground can occur above that limit. In our original setup this happened immediately after reaching 600~V. We decided to remove the anode cable and solder the HV-connector directly onto the anode wire outlet, as can be seen in the picture on the right hand side of Fig.~\ref{Fig04}. This also reduces the noise level and increases the simplicity of the detector. The quality of the cathode cable does not impact to the operation of the detector.

\section{Experimental setup}\label{sec:setup}

The detector was flushed over-night with a P10 gas mixture (Ar 90~\% / methane 10~\%) to remove oxygen, which is highly electronegative and captures some of the primary electrons from the ionization process. Even though some electrons will reach the avalanche region close to the anode, the loss of primary electrons would result in a loss of the proportional character of the counter. The gas flow during the night was set to $\approx 20$~ml/min. The gas outlet of the detector is either vented into the atmosphere or connected to a gas-recycling circuit, if available. Additional beverage can detectors can be connecting in series, but the flushing-time will increase with each added device. If the gas is vented into the atmosphere, the gas outlet tube should be at least a couple of centimeters long to prevent back-flow diffusion. 

\begin{figure}[b]\centering
\includegraphics[width=0.45\textwidth]{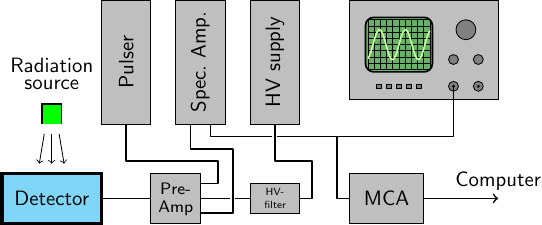}
\caption{Schematics of the measurement setup.\label{Fig05}}
\end{figure}

\begin{figure}[b]\centering
\includegraphics[width=0.35\textwidth]{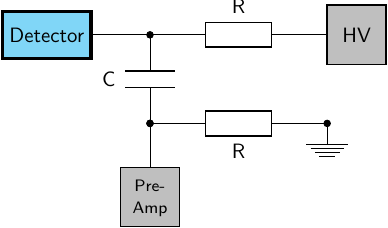}
\caption{Circuit of the capacitive decoupling of HV and signal.\label{Fig06}}
\end{figure}

For the measurement an analog oscilloscope (HP COS 6100, 100~MHz), a pre-amplifier (Ortec Pre Amp 142 IH), a linear spectrum amplifier (Ortec 855 Dual Spec Amp), a pulse generator (BNC Pulse Generator Model BL-2), a high voltage supply ($\Sigma$ Silena Milano), and a multi-channel-analyzer (Pocket MCA, AmpTek) were used.  The experimental setup is shown schematically in Fig.\,\ref{Fig05}.
The Ortec pre-amplifier uses capacitive decoupling to separate the signal from the HV.   If the pre-amplifier does not have an integrated capacitive decoupling of HV and signal it is easy to build a separate decoupling box using the circuit shown in Fig.\,\ref{Fig06}.
The linear amplifier provided gains between 5 and 1200. The pulse generator was used to calibrate the electronics. The HV supply can provide up to $5$~kV with a current limit of 1~mA.
The signal from the linear amplifier was read out in parallel by the analog
oscilloscope and the multi-channel-analyzer which was interfaced with a computer. 

An essential part of the operational preparations is noise reduction. Standard aluminium kitchen foil proved to be highly effective for shielding. For the noise study, the HV was slowly ramped up to values well above 1~kV, while the gas flow through the detector was kept constant at $\approx 60$~ml/min of P10. The noise was observed with the oscilloscope and aluminium foil was wrapped around the end-caps of the detector. A firm contact to the cathode (ground) was obtained with self-adhesive conducting copper-tape. This reduced the noise to amplitudes  below 100~mV for a gain of 120. 

\begin{figure}[]
\includegraphics[width=0.32\textwidth]{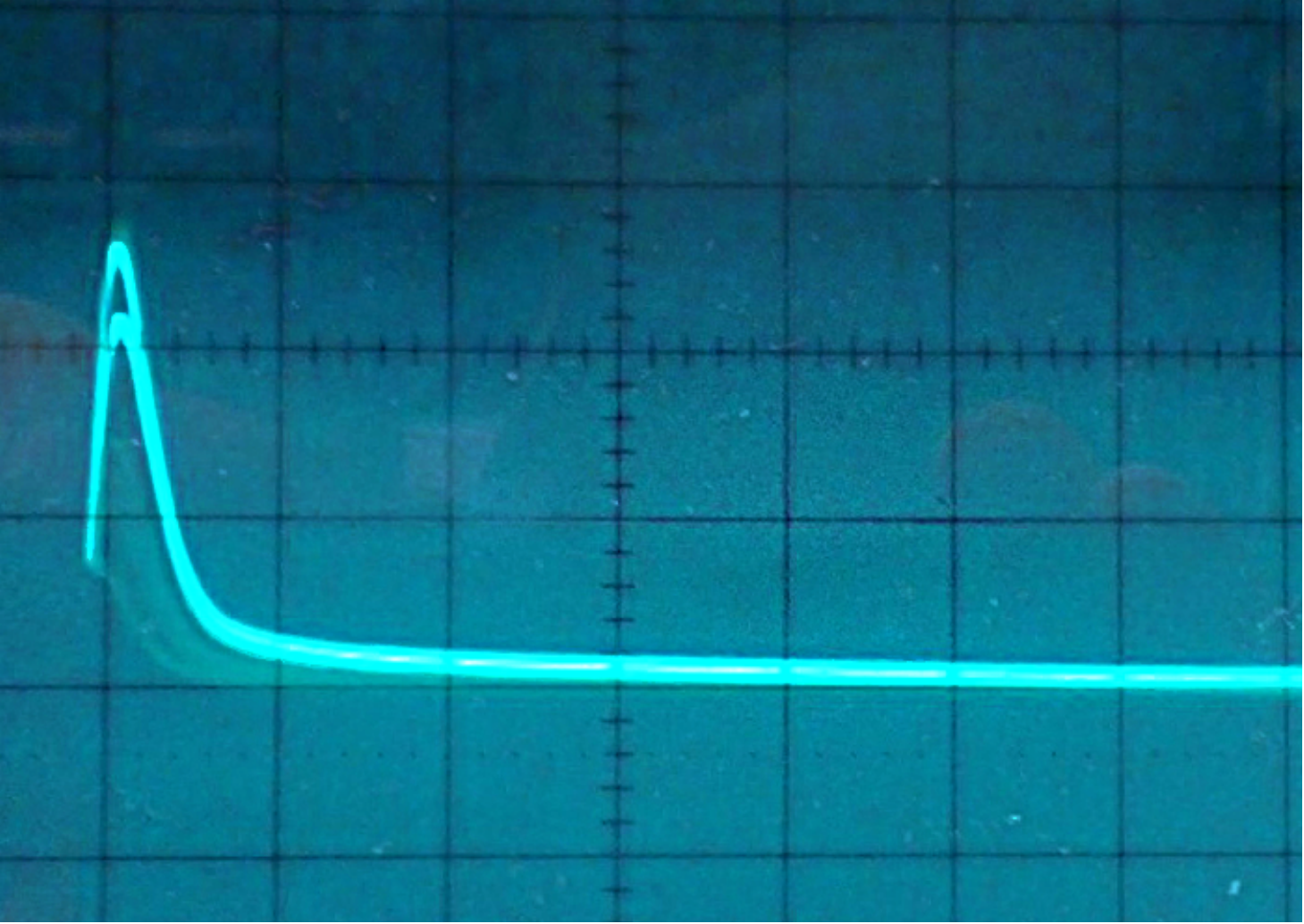}
\caption{Detector signals seen on the oscilloscope when a $^{55}$Fe source was put in front of the detector. The horizontal and vertical scales are 50~$\mu$s/division and 100~mV/division, respectively. The shaping time and gain of the amplifier used in this measurement was 3~$\mu$s and 120, respectively. \label{Fig07}} 
\end{figure}

We used $^{55}$Fe and $^{241}$Am isotopes to test the detector. The latter can be obtained from
commercial ionizing smoke detectors.\cite{smokedetector} For our measurements, the sources were put on top of the detector. A simple lead collimator can be used to reduce signal pileup\cite{collimator,pileup} if the source activity is too high. An example of a typical detector signal is shown in Fig.\,\ref{Fig07}. 

The analysis of the measured data can be done manually, with spreadsheet programs, or with scientific software such as \textsc{root}, a freely available 
object-oriented data analysis framework\cite{root} that was used in our case.

\section{Electronics calibration}\label{sec:electrocalib}

To calibrate the electronics a negative test signal from the pulse generator was sent to the pre-amplifier and the oscilloscope. The test signal was a tail pulse with a steep falling edge and a slow rise. A negative pulse was used since the signal from the anode wire of the proportional counter is negative. The amplitude of the test pulse was measured with the oscilloscope. The response of the electronic chain was recorded with the MCA. A gain factor of 120 was used during the calibration process. 

During the electronics calibrations, the detector was connected to the pre-amplifier and the bias voltage was set to zero. 
For accurate calibrations, all devices that are used for the measurements should be connected and turned on. 
The response of the electronics to test pulses with amplitudes in the order of 100~mV was measured and the results are shown in Fig.\:\ref{Fig08}.  Fig.\:\ref{Fig08}~(bottom) shows the MCA channel histograms for the different test pulses. The test capacitance of 1~pF was taken into account to calculate the injected charge  $Q=C\,V$, shown by the data points in Fig.\:\ref{Fig08}~(top). A linear function was used to fit the data and obtain the following relation between the charge $Q$ and the MCA channel number:
\begin{figure}[b]\centering
\includegraphics[width=0.45\textwidth]{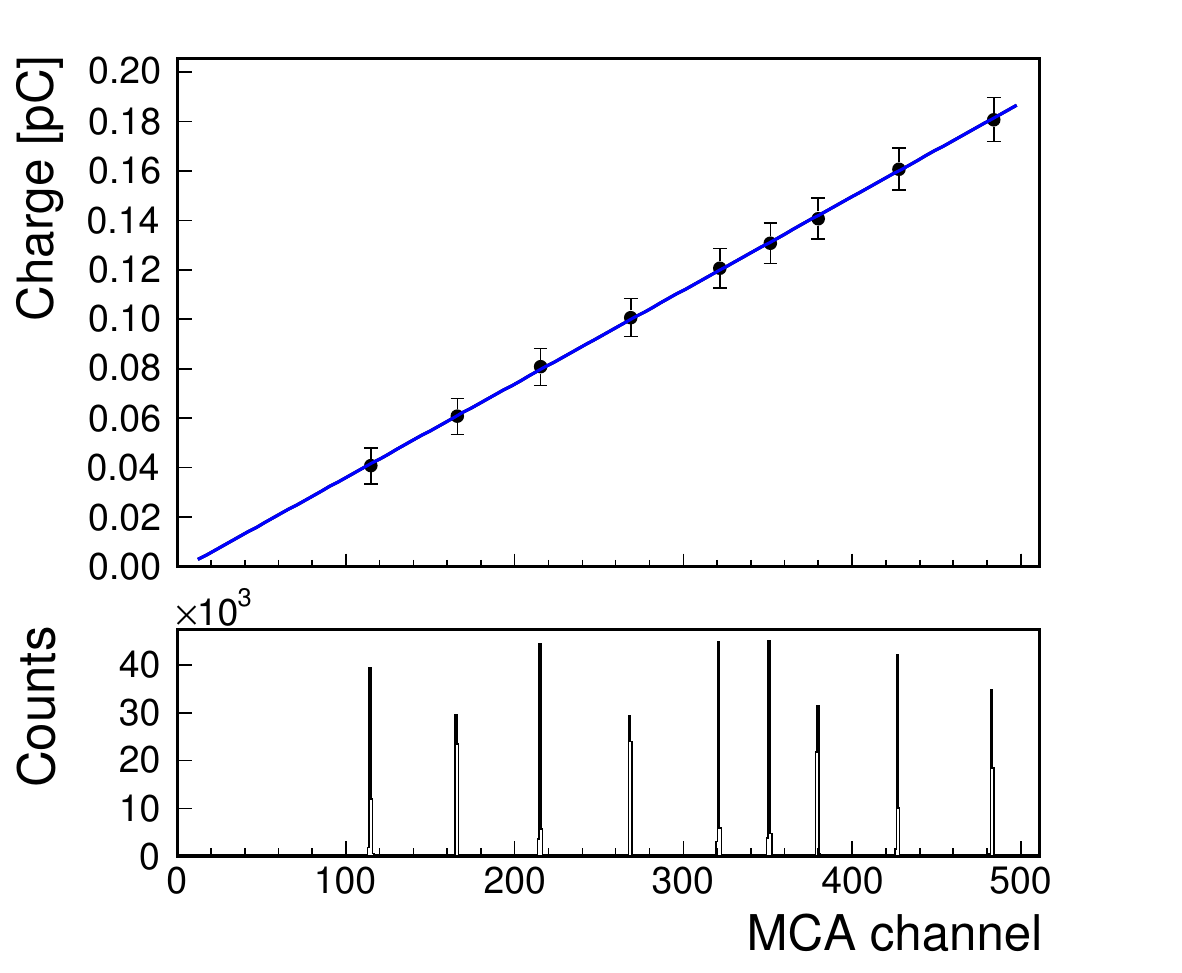}
\caption{Calibration plot of the readout electronics (top) and the recorded MCA spectrum(bottom).\label{Fig08}} 
\end{figure}
\begin{equation}
\mathrm{Charge}=g\cdot\mathrm{Channel_{\,MCA}}+h,
\end{equation}
where $g=3.8\times10^{-4}\pm2.3\times10^{-5}$~pC/channel and $h=-1.9\times10^{-3}\pm7.1\times10^{-3}$~pC. The uncertainty in the calibration constants is dominated by the uncertainty in the measurement of amplitude of the test pulse, estimated by eye to be 5~mV. 
This calibration will be used to probe the detectors operational region.

\section{Measurement of $X$ and $\gamma$-ray spectra}\label{sec:spectra}

\begin{figure}[]\centering
\vspace{-0.4cm}\includegraphics[width=0.45\textwidth]{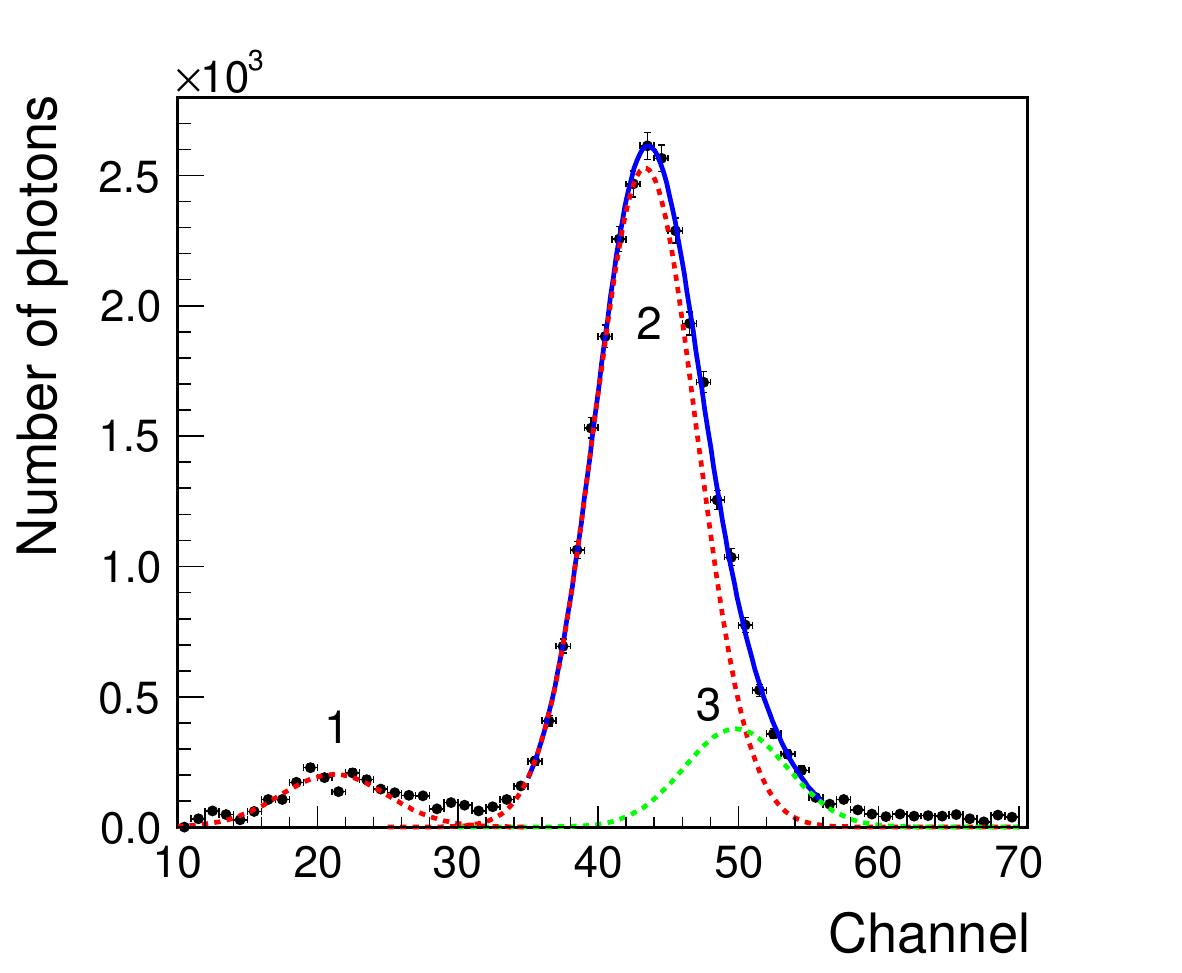}
\caption{Energy spectrum obtained with $^{55}$Fe. 
\label{Fig09}} 
\end{figure}

\begin{figure}[]\centering
\includegraphics[width=0.45\textwidth]{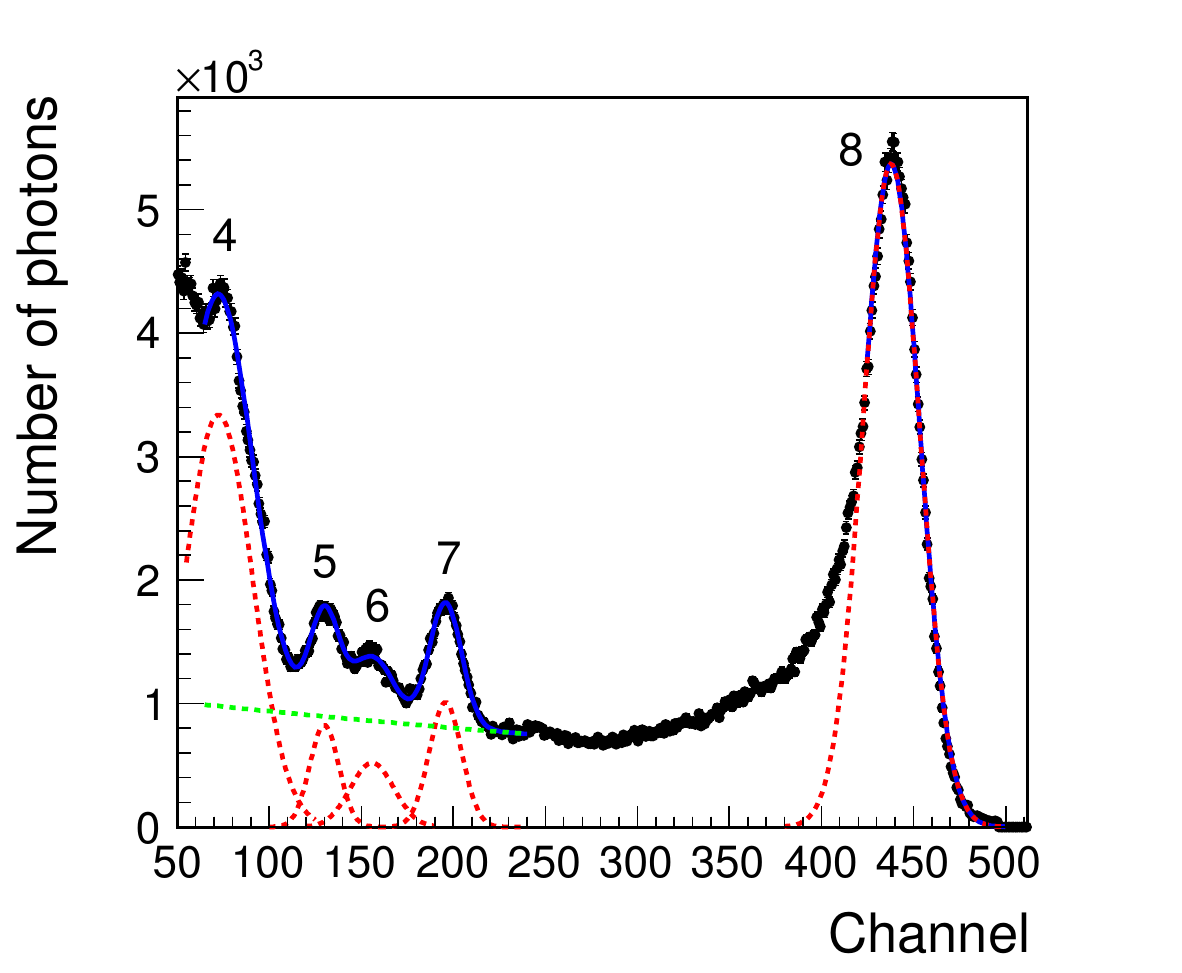}
\caption{Energy spectrum obtained with $^{241}$Am. 
\label{Fig10}} 
\end{figure}

The bias voltage was ramped up until a clear signal was seen on the oscilloscope when the $^{241}$Am source was placed on top of the detector. The HV and the amplifier gain were adjusted to observe the whole spectrum of $^{241}$Am over the full range of the MCA. At a voltage of $2.25$~kV and a gain of $120$, the spectra of both isotopes were recorded. In addition, a background spectrum was acquired to be able to do a background subtraction.\cite{background}  The energy spectra obtained with $^{55}$Fe and $^{241}$Am are shown in Figs.~\ref{Fig09} and \ref{Fig10}, respectively.

For each spectrum, the most prominent peaks were used to calibrate the energy scale of the measurement. From common isotope tables\cite{isotopetables} and the table of nuclides\cite{AmDecayScheme, FeDecayScheme} it is well known that the decay of $^{55}$Fe is accompanied by prominent $X$-ray emissions originating from the relaxation process of the $^{55}$Mn daughter. These $X$-ray emissions, with an energy of $5.89(875)$~keV (16.2~\%, $K_{\alpha_1}$), $5.88(765)$~keV (8.1~\%, $K_{\alpha_2}$) and $6.49(045)$~keV (2.9~\%, $K_{\beta_{1/3}}$), are the dominating spectrum of $^{55}$Fe. The energy resolution \cite{energyresolution} of the detector is not good enough to separate these emissions\cite{fsauli} due to the large gas volume of the can and the in-homogeneity of the anode wire that influences the local electric field. 

An attempt was made to fit a double Gaussian distribution function to the visible peak, as shown in Fig.\,\ref{Fig09}. The two almost not distinguishable $K_{\alpha}$ emissions are represented by fit number 2 and the $K_{\beta}$ emissions by fit number 3. The intensity ratio of the sum of $K_{\beta_{1/3}}$ and $K_{\alpha_{1/2}}$ is approximately 12\%. This ratio and a similar width of the two Gaussian distributions were used to constrain the fit. The resulting double Gaussian fit converged with a $\chi^2$ value of 1.29.

\begin{figure}[]\centering
\includegraphics[width=0.45\textwidth]{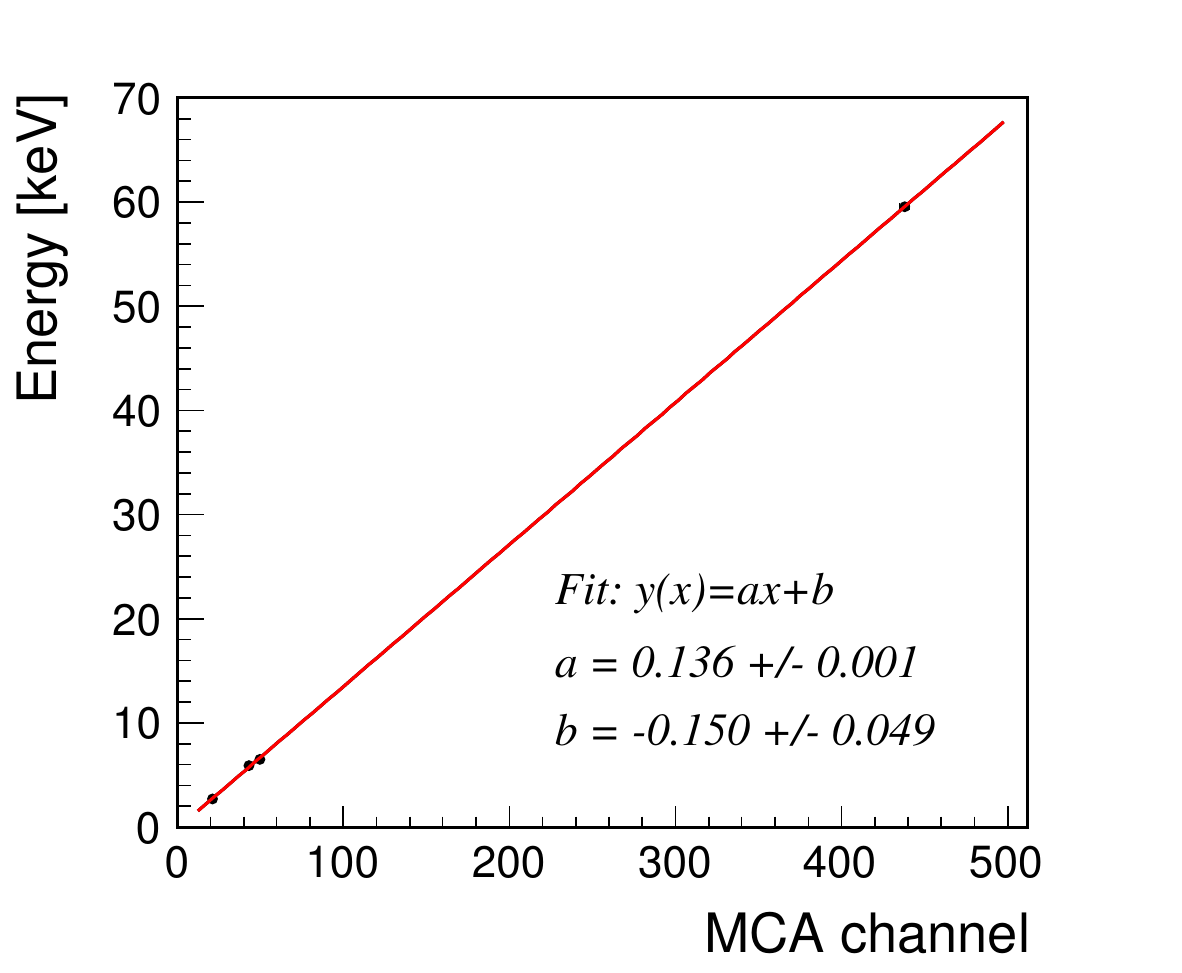}
\caption{Energy calibration of the MCA. 
\label{Fig11}} 
\end{figure}

The most prominent escape peak of argon (lowered by $3.19(199)$~keV; peak 1 in Fig.\,\ref{Fig09}) was also fitted with a Gaussian function. Escape peaks are due to secondary $X$-ray emissions, originating from the detector gas, that escape from the detector.
In the energy spectrum obtained with $^{241}$Am, shown in Fig.\ref{Fig10}, the most prominent emission line is the $59.54$~keV line due to a transition in $^{237}$Np (peak 8 in Fig.\,\ref{Fig10}.) The peak was fitted with a single Gaussian function. 
Using those known peaks of $^{55}$Fe and $^{241}$Am an energy calibration was performed with the result shown in Fig.\,\ref{Fig11}.

Four other visible peaks of the $^{241}$Am spectrum, between channel 50 and 220 in Fig.\,\ref{Fig10}, were fit with Gaussian distributions and an exponential background. 
The details of the peaks used for the energy calibration as well as details of the four other ``unknown'' peaks in the $^{241}$Am spectrum are listed in Table~\ref{Tab02}.
\begin{table}[]\centering
\caption{Details of the peaks observed in the spectra obtained with $^{55}$Fe and $^{241}$Am. Peaks are numbered with increasing
energy. The systematic and statistical uncertainties are added in quadrature.\label{Tab02}}
\begin{ruledtabular}
\begin{tabular}{ l  c  c  c}
Peak & Centroid [channels] & Energy [keV] & FWHM [keV]\\
\hline               
1 & 21.2 $\pm$ 0.4 & 3.2 (calib.) & 0.97 $\pm$ 0.15\\%
2 & 43.4 $\pm$ 0.3 & 5.9 (calib.) & 0.82 $\pm$ 0.12\\%
3 & 49.8 $\pm$ 0.3 & 6.49 (calib.) & 0.82 $\pm$ 0.13\\%
4 & 72.4 $\pm$ 0.6 & 9.72 $\pm$ 0.12 & 5.59 $\pm$ 0.23\\%
5 & 129.9 $\pm$ 1.0 & 17.56 $\pm$ 0.19 & 2.18 $\pm$ 0.16\\%
6 & 155.9 $\pm$ 1.2 & 21.10 $\pm$ 0.23 & 3.45 $\pm$ 0.26\\%
7 & 195.8 $\pm$ 1.4 & 26.53 $\pm$ 0.27 & 2.47 $\pm$ 0.13\\%
8 & 437.9 $\pm$ 3.1 & 59.54 (calib.) & 4.60 $\pm$ 0.12\\%
\end{tabular}
\end{ruledtabular}
\end{table}
Using the calibration function shown in Fig.\,\ref{Fig11}, the energies and widths of the ``unknown'' peaks were determined.

\subsection*{Uncertainties}

Apart from the statistical uncertainties, systematic uncertainties were also taken into account and standard textbook error propagation techniques were used.\cite{cowan} 

Systematic errors are associated with the HV power supply, the amplification chain, and the MCA. The instability of the Silena Milano power supply was estimated to be 1~V.  The corresponding  uncertainty in terms of MCA channels is estimated by using the result of the operation region test (see Fig.~\ref{Fig12}). The uncertainties of the pre-amplifier and the spectrum amplifier were taken from the specification sheet, each contributing a integral nonlinearity of $<\pm 0.05\;\%$. The MCA has a differential nonlinearity of 0.6~\%  and an integrated nonlinearity of 0.02~\%.   Errors resulting from pressure and temperature changes of the environment during the measurement as well as uncertainties due to charge-up and ion build-up were not studied. 
The total systematic uncertainty was added to the statistical uncertainty by quadrature and shown in Table~\ref{Tab02}.

\subsection*{Discussion of the results}

Peaks 1, 2, 3 and 8, listed in Table~\ref{Tab02}, were used to obtain the energy calibration. Peak number 1 with an energy resolution of 50~\% is not well pronounced due to the large gas volume and the non-ideal background subtraction. The characteristic $^{55}$Fe peak was fit with a double Gaussian distribution (peaks 2 and 3). The energy resolution is $\approx$~27~\% for the 5.9~keV emission and $\approx$~36~\% for the 6.49~keV emission.   Peak 8 is 
the characteristic $^{241}$Am peak at 59.54~keV which has an energy resolution of about 
9~\%. This peak shows a non-Gaussian tail on the left side, mostly caused by photo-electrons escaping the gas volume before losing all their energy in ionizing collisions. Peaks~5 and 7 are well-pronounced peaks with similar energy resolutions of about 16~\% and about 12~\%, respectively. Peak~7 appears to correspond to the $26.34$~keV transition in $^{237}$Np with an intensity of 2.4~\%. The origin of peak number 5 is less clear. The closest emission found in radiation tables of the $^{241}$Am \cite{isotopetables,AmDecayScheme} is the $17.75$~keV $X$-ray emission from the $L_{\beta1}$ shell in $^{237}$Np with an intensity of 5.7~\%. Other $X$-ray emissions from the $L_{\beta}$ shells also contribute. A weighted average of those suggests a peak centroid of 17.48 which is well below one standard deviation off our measured peak.
For peak number 6 no direct candidate was found. Peak 6 might be the result of overlapping $X$-ray emissions from the $L_{\gamma}$ shells. The weighted centroid of the $L_{\gamma}$ shells $X$-ray emissions suggests a peak at 21.04~keV that is compatible with our measured peak. 
Another $X$-ray emission from $L_{\alpha1}$ at 13.94~keV with an intensity of 9.6~\% is expected. Our recorded spectrum reveals a possible peak around 10~keV (peak 4), but no clear evidence is found without applying a thorough background subtraction. 

It should be mentioned that the detection efficiencies for the transitions in the energy spectra are energy dependent. As shown in Fig.\,\ref{Fig03}, low-energy photons have a higher attenuation in the aluminium can wall than high-energy photons. The effect of attenuation can be corrected using the efficiency distribution shown in Fig.\,\ref{Fig03}.

\section{Test of the operating region}\label{sec:hvscan}

To reveal the full potential of the detector, the operational HV region was scanned in order to understand its proportionality behavior and measure the gas gain. The idea was to measure the prominent characteristic peaks of the two isotopes over the full operational voltage range available.\cite{propcountrange} The location of the peaks observed with the MCA were converted into induced charge on the anode wire, using the electronics calibration function shown in Fig.\,\ref{Fig08}.

The measurement was started with $^{55}$Fe and the maximum gain of the amplification system (factor 1200). The bias voltage was set to the minimum level of $1.7$~kV at which a clear signal could still be seen with the oscilloscope and recorded at the lower end of the MCA scale. A recording time of $\approx$~200~s was sufficient to acquire just enough statistics to resolve the characteristic peaks and short enough to finish the measurement in reasonable time. We recorded data for voltage steps of 50 to 100~V until we reached the minimum gain of the amplification chain and the upper limit of the MCA range.

Next, $^{241}$Am was used with its characteristic $\gamma$-ray emission at a ten times higher energy than the characteristic $^{55}$Fe $X$-ray emission. This allowed us not only to extend the scan of the operational range but also to show the proportional character of the detector. The observed charge curve obtained with $^{241}$Am should be shifted upwards and be parallel to the curve obtained with $^{55}$Fe. Again, spectra were recorded in steps of 50 or 100~V over the whole amplification range. The measurement started with the maximum gain and a bias voltage of $1.1$~kV, at which one still was able to identify the peak signal on the oscilloscope. 

\begin{figure}[]\centering
\includegraphics[width=0.44\textwidth]{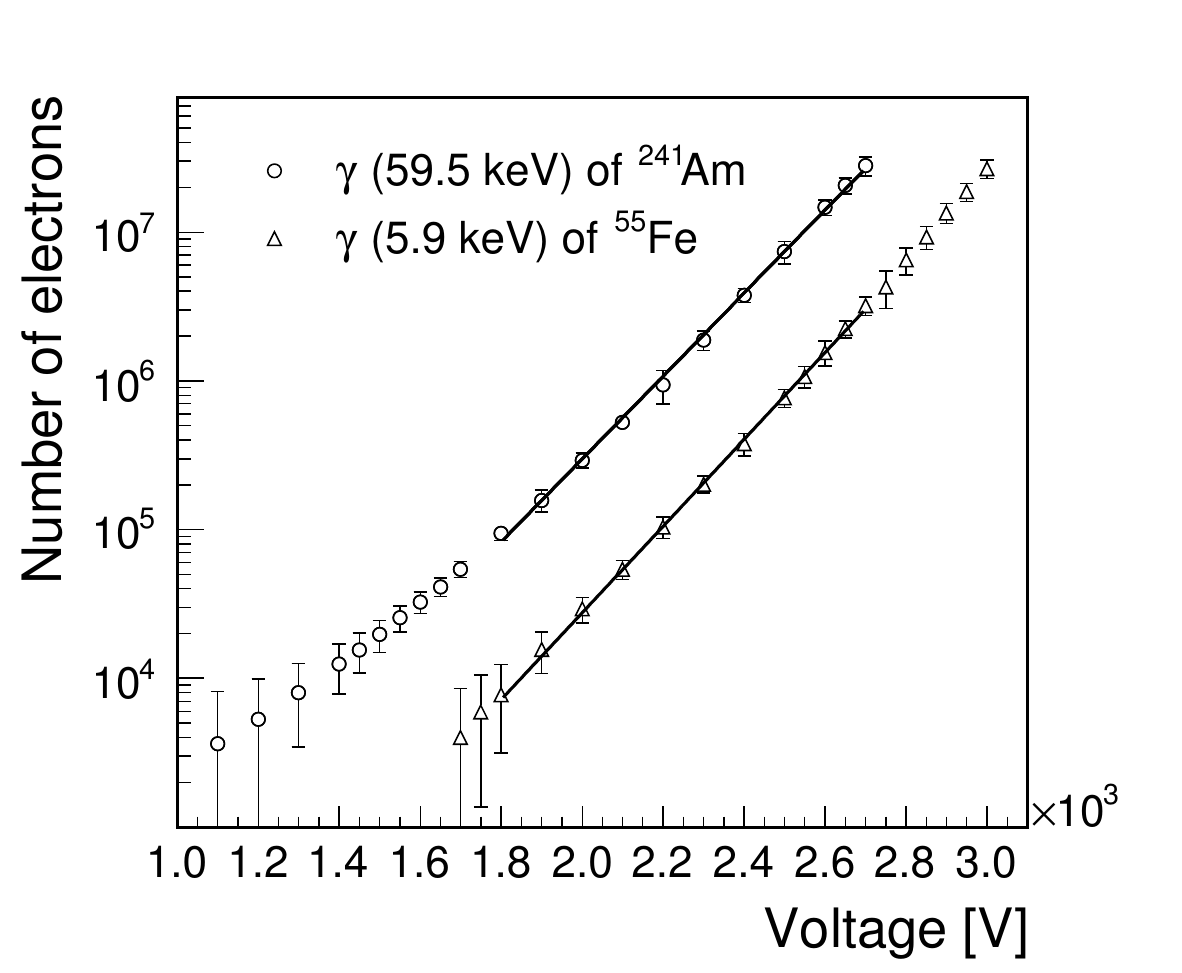}
\caption{Measured charge versus the applied bias voltage for measurements with $^{55}$Fe and $^{241}$Am. Error bars include statistical and systematic uncertainties.\label{Fig12}}
\end{figure}

For all the recorded spectra the characteristic peaks and their details were extracted. The centroids of the peaks were converted into total charge or the number readout electrons using the measured electronics calibration function. These results are shown in Fig.\,\ref{Fig12}. Exponential fits were carried out for both data sets with the following function.
\begin{equation}
N_{e^-}=\exp{\left(s\cdot U + t\right)}.
\end{equation}
The measured slope was $s=6.70\pm0.23\:(6.42\pm0.15)\times 10^{-3}$~V$^{-1}$ and the offset was \mbox{$t=-3.18\pm0.55\:(-0.23\pm0.34)$} for the $^{55}$Fe ($^{241}$Am) data.
In addition to the uncertainties already discussed, the errors include the effect of a 5~V uncertainty of the manual setup of the HV power supply. 

\begin{figure}[]\centering
\includegraphics[width=0.44\textwidth]{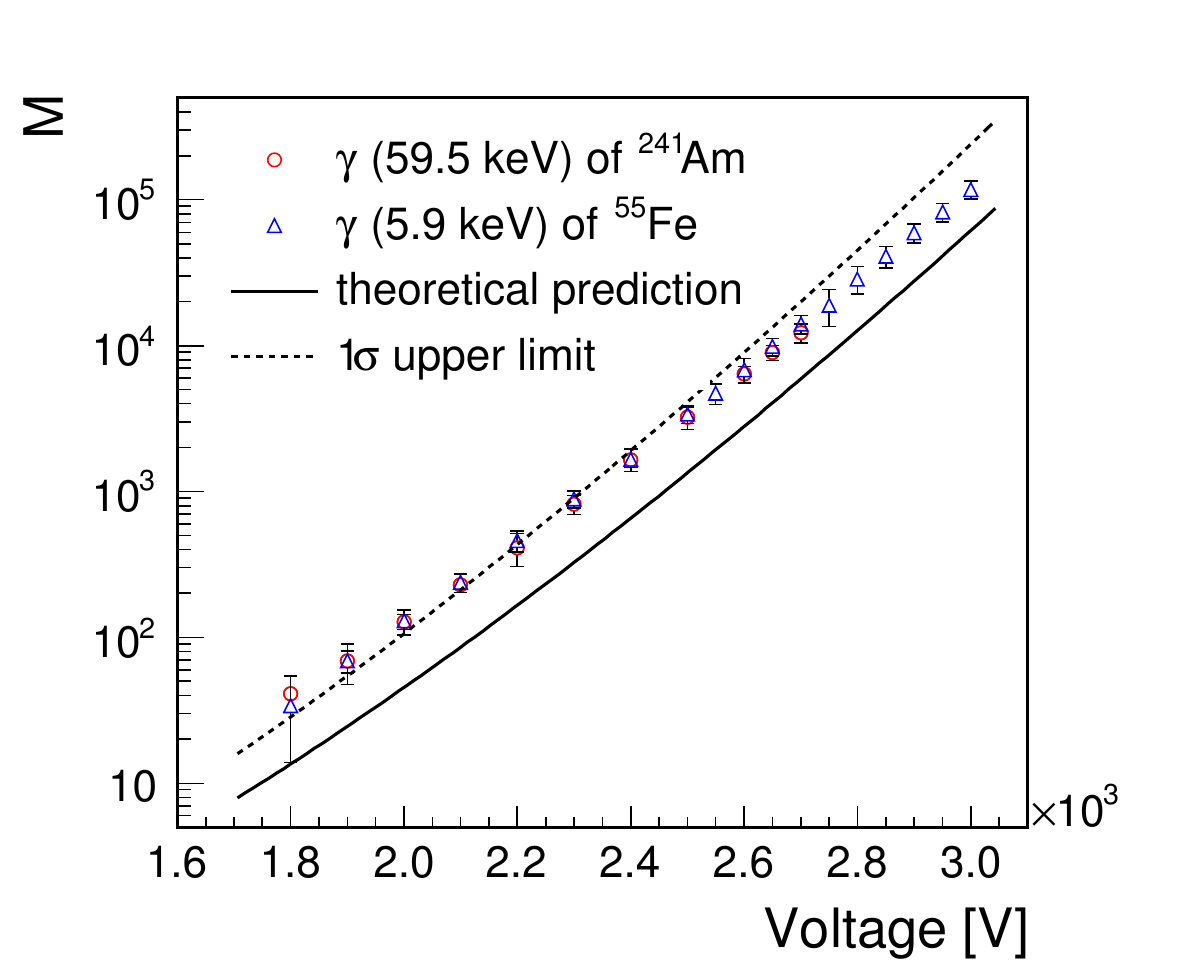}
\caption{Gas multiplication factor $M$ versus bias voltage. The solid line shows the theoretical prediction obtained using the Diethorn formula.\cite{wolffi}\label{Fig13}}
\end{figure}

The proportional character of the detector is consistent with the data shown in Fig.\,\ref{Fig12}. Plotted on a logarithmic scale, the proportional operational region is the linear part of the curves. The curves shown in Fig.\,\ref{Fig12}. are almost parallel and shifted by a factor close to 10, which is expected as the characteristic isotope peaks are a factor of 10 apart in terms of energy. 
With about 26~eV needed to create an electron-ion pair in argon gas, we can estimate the gas multiplication factor knowing the energy of the incident $X$- or $\gamma$-rays. We assumed that all the energy was absorbed by the gas, which is not totally true for the 59.54~keV $\gamma$-ray from $^{241}$Am. The gas multiplication factor $M$ was calculated by using the formula,
\begin{equation}
M=\frac{Q}{n_0e^-}=\frac{Q}{(E_{\mathrm{rad}}/E_{\mathrm{ionpair}})e^-},
\end{equation}
where $Q$ is the collected charge, $E_\text{rad}$ is the energy of the incident radiation, and $E_\text{ionpair}$ is the energy required to create an electron-ion pair. The 5.9~keV $X$-ray of $^{55}$Fe produces 227 electron-ion pairs and the 59.54~keV $\gamma$-ray of $^{241}$Am produces 2,290 electron-ion pairs. The gas multiplication factors are shown in Fig.\,\ref{Fig13} and are in satisfactory agreement with theoretical predictions.
The predictions were obtained using the Diethorn formula,\cite{wolffi}
\begin{equation}
\ln M = \frac{V}{\ln(b/a)}\frac{\ln 2}{\Delta V}\left[\ln \frac{V}{pa\ln(b/a)}-\ln K\right],
\end{equation}
where $V$ is the bias voltage, $a$ and $b$ are the anode and cathode radii, and $p$ the gas pressure. The parameters $K$ and $\Delta V$ depend on the properties of the gas mixture.\cite{wolffi}

The Geiger-M\"uller region is unfortunately not accessible due to the argon-methane gas mixture that was used. Even though the data recording was stopped at 3~kV, the HV was ramped up to 4.7~kV out of curiosity. Constant discharge was observed, including a glow around the anode wire, and sparks within the gas volume were observed through the Acrylglass end-cap window. Afterwards, the detector was not able to measure any radiation even at voltages around 1.4~kV. After flushing the detector over night it returned to a state, able to detect radiation at the nominal level. 
Recombination of positive ions at the cathode might have been slowed down by the polymer coating that was not removed. Plasma discharge, which, produces free radicals that attach to the anode-wire, result in detector aging.

\section{Conclusions}\label{sec:conclusions}

We demonstrated that it is relatively easy and inexpensive to build a fully functioning gaseous proportional counter for  measuring radioactive $X$- and $\gamma$-ray emissions. 
Building such a device offers great educational benefits for students at all levels. Not only do the students learn how to build an inexpensive proportional counter from scratch, they also are introduced to high voltage systems and electronic readout devices used in actual research. In addition to the experimental techniques, students may also gain knowledge in analyzing the recorded data and discussing experimental uncertainties.

Overall the detector is fun to build, and the results are astonishing, especially in
the context of simplicity and inexpensiveness.

\begin{acknowledgments}
We thank the staff of the Detector Laboratory of Helsinki Institute of Physics and University of Helsinki for their valuable help and discussions, i.e. Eija Tuominen, Jouni Heino, Rauno Lauhakangas and Raimo Turpeinen. We also thank Ismo Koponen and Michael Albrow for reading our manuscript.
\end{acknowledgments}


\begin{thebibliography}{5}
\bibitem{cansize} A 0.5~l can proved to be better than a 0.33~l can due the higher length to width ratio for the larger can. It results in a more homogeneous electric field.
\bibitem{collimator} A collimator is not needed if weak sources are used, such as usually available in high-schools.
\bibitem{pileup} Due to restricted time resolution it is called pileup if more then one photon contributes to the signal.
\bibitem{background} For best results a background spectrum should be taken before and after every measurement.
\bibitem{energyresolution} Energy resolution in relative terms is usually defined as the full width at half maximum of the recorded spectrum peak centroid divided by the energy of the emission line.
\bibitem{propcountrange} As can be seen in Fig.~50 of reference \cite{sauli}.
\end{thebibliography}
\end{document}